\documentclass[doublecol]{epl2}
\usepackage{soul}
\title{Created-by-current states in long Josephson junctions}
\shorttitle{Created-by-current states in long Josephson junctions} 

\author{T.L. Boyadjiev\inst{1,2} \and O.Yu. Andreeva\inst{3} \and E.G. Semerdjieva\inst{1,4} \and Yu.M. Shukrinov\inst{1,5}}
\shortauthor{T.L. Boyadjiev \etal}

\institute{
  \inst{1}  JINR, Dubna, Moscow Region, 141980, Russia\\
  \inst{2}  Sofia University, Sofia, 1164, Bulgaria\\
  \inst{3}  Tumen Thermal Networks OAO TGK 10, Tobolsk, 626150, Russia\\
  \inst{4} University of Plovdiv, Plovdiv, 4000, Bulgaria\\
  \inst{5} Physical Technical Institute, Dushanbe, 734063, Tajikistan}

\abstract{ Critical curves "critical current - external magnetic field" of long Josephson junctions with
inhomogeneity and variable width are studied. We demonstrate the existence of the regions of magnetic field
where some fluxon states are stable only, if the external current through the junction is different from zero.
Position and size of such regions depend on length of the junction, its geometry, parameters of inhomogeneity
and form of the junction.  The noncentral (left and right) pure fluxon states are appeared in the inhomogeneous
Josephson junction   with increase in the junction length. We demonstrate new bifurcation points with change in
width of the inhomogeneity and   amplitude of the Josephson current through the inhomogeneity.}

\begin{document}

\maketitle

\section{Introduction} Starting from the classical paper of Owen and Scalapino,~\cite{os_67} the long
Josephson junctions (LJJ) are attracting attention both experimentalists and theoreticians. They are good
candidates for a wide range of applications such as superconducting quantum interference devices, Josephson
voltage standards, logic elements, and Josephson flux-flow oscillators because of their nonlinear behavior and
quantum effects. ~\cite{bp_84,kkl_86,ped97} From the other side, the nonlinearity makes the LJJ a very complex
system, and some aspects of  its properties are not investigated up to now. Particularly, the experimental
methods to determine the vortex structure under the critical curve are not developed well yet. But there are
different kind of stable states in the LJJ, which bifurcation curves are close to the critical curve of the
junction and manifest themselves in the experiments.~\cite{bp_84,vdks_88} Particularly, some "metastable" states
(experimental points under overlap curve) were observed in Ref.~\cite{schwidtal70}. Static bound states of
fluxons in LJJ with artificial inhomogeneities were experimentally found in Ref.~\cite{vdks_88}. Experimental
and theoretical investigation of such states is an interesting and actual problem.

An important step of the sample characterization in the experiment with LJJ's is the measurement of its static
properties, i.e. the dependence of the critical current $I_c$ on external magnetic field $H$. This dependence
provides a technique to evaluate the several important parameters of LJJ from the experiment, particularly, a
critical current density, a magnetic flux penetration field, an effective magnetic
thickness.~\cite{gold-current99} Since the solution of the corresponding boundary value problem (BVP) cannot be
performed analytically in general case, the most straightforward way to study the dependence $I_c(H)$ is its
numerical simulation.

In our previous study we investigated the stability of the vortices in LJJ with inhomogeneities
(LJJI)~\cite{fgbp_87,review_07} and vortex structure in exponentially shaped Josephson junctions
(ESJJ).~\cite{sbs_04,ssb_05,sbs_05} The obtained bifurcation curves of the mixed fluxon-antifluxon states in
LJJI~\cite{review_07} and pure fluxon states in ESJJ demonstrate the intervals of magnetic field, where these
states are stable only, if the current through the junction is not equal to zero. These peculiarities were not
mention at that time.  We call these states as the CbC-states (created-by-current), and the corresponding
regions of the external magnetic field as CbC-regions.

In this Letter we study the CbC-states  of mixed fluxon states $\Phi ^{-1}\Phi ^{1}$, $\Phi ^{1}\Phi ^{-1}$ and
pure fluxon states $\Phi ^{n}$.  We demonstrate new bifurcation points with change in width of the inhomogeneity
and amplitude of the Josephson current through the inhomogeneity. The influence of the model parameters on CbC
regions in the LJJI and ESJJ is investigated.
\section{Method of calculation}First we start with LJJI case. We consider the Josephson junctions in the overlap geometry.~\cite{bp_84} In
order to study a linear stability of the static distribution $\varphi(x)$ of magnetic flux $\phi(t,x)$, we write
the perturbed expression for $\phi(t,x)$ as $\phi(t,x) = \varphi(x) + \varepsilon \xi(t) \psi(x)$, where
$\varepsilon$ is a parameter ~\cite{ftbk_77,galfil_84}. Then in a first approximation with respect to
$\varepsilon$, we get the following boundary value problem (BVP) for $\varphi(x)$
\begin{equation}
\label{steq}  -\varphi_{xx} + j_C(x) \sin\varphi + \gamma = 0, \hspace{0.2cm} \varphi(-l)= \varphi(l)=h_e
\end{equation}
and  the corresponding Sturm-Liouville problem (SLP)
\begin{equation}
\label{slp}
 -\psi_{xx} + \sigma(x) \psi_x + q(x) \psi = \lambda \psi \hspace{0.3cm} \psi_x(-l)=\psi_x(l)=0
\end{equation}
Here all the quantities are in dimensionless form ~\cite{kkl_86}, $h_e$ is the external magnetic field, $\gamma$
is  the external current, and $\lambda$ is an eigenvalue of the SLP, $2l$ is a length of the junction. The
function $j_C(x)$ represents the Josephson current amplitude, so $|\,j_C(x)| \le 1$. Because of the nonlinearity
the BVP, eq.(~\ref{steq}) has more then one solution for given parameters (counted set in case $h_e = 0$,
$\gamma = 0$, $l \to \infty$)~\cite{ikk_94}.  The potential $q(x) = j_D(x) \cos\varphi(x)$ is defined by the
concrete static solution $\varphi(x)$, which is asymptotically stable with respect to small perturbations, if
$\mathop {\lim }\limits_{t \to \infty } \left|\, \xi(t) \right| \to 0$. As the junction's length $l < \infty$ in
case under consideration and the potential $|q(x)| \le 1$, the SLP eq.(~\ref{slp}) is regular, so there exists a
discrete low-bounded set of real eigenvalues $\lambda_n$, $n = 0, 1, 2,\ldots$  and the minimal one $\lambda_0
\ge -1$.~\cite{ls_75} The corresponding eigenfunctions satisfy the norm condition $ \int\limits_{-l}^l
\psi^2(x)\,dx - 1 = 0$. In case $\lambda_0 < 0$ the static solution $\varphi(x)$ is unstable, i.e. $|\,\xi(t)|
\to \infty$, when $t\to\infty$. Solutions of the BVP eq.(~\ref{steq}), eq.(~\ref{slp}) describe the different
fluxon states in LJJ.  Numerical simulation simplifies the study and makes it possible to estimate the range of
variation of the parameters in which one can expect stability or instability of the magnetic flux distributions
in the Josephson junction.

The models of LJJ depend on $m$ parameters, such as junction's length $2l$, external magnetic field $h_e$,
external current $\gamma$ and so on. Let us denote the vector of all the parameters by $p$. In all cases we
presume that the possible solutions of non-linear BVP eq.(~\ref{steq}) continuously depend on the set $p$, i.e.
$\varphi = \varphi(x,p)$. It follows that the potential of SLP eq.(~\ref{slp}) generated by the solution
$\varphi(x,p)$, depends on the model parameters $p$ as well, i.e., $q = q(x,p)$. Finally, the corresponding
eigenvalues and eigenfunctions of eq.(~\ref{slp}) are continuous functions of $p$, i.e. $\lambda_n =
\lambda_n(p)$, $\psi_n = \psi_n(x,p)$.

Hence, the static solution $\varphi(x,p)$ is asymptotically (exponentially) stable with respect to small
space-time perturbations in some bounded subset of the parameter region, if the minimal eigenvalue satisfies
$\lambda_{0}(p) > 0$.~\cite{galfil_84} Every point  on the hypersurface $\lambda_{0}(p) = 0$ is the bifurcation
point for the solution under consideration. The values of the parameters which satisfy $\lambda_{0}(p) = 0$, are
called the bifurcation (critical) values for this solution. The cross section of this hypersurface  by a
hyperplane, which correspond to fixed values of $m-2$ parameters, determine the bifurcation curve for the rest
two parameters. From the experimental point of view, the most important are the bifurcation curves of kind
``critical current-magnetic field''
\begin{equation}
\label{gamhe}
\lambda_{0}(\gamma, h_e) = 0\,.
\end{equation}
The numerical algorithm for the determination of the bifurcation curves is proposed in ~\cite{bpp_jinr88} and
was successfully applied to various physical problems ~\cite{review_07}
\section{Bifurcation curves in LJJ with inhomogeneity}The static configurations of finite-length junction in the presence of an external magnetic field were studied
by many authors.~\cite{os_67,fgbp_87,pagano,cfgsv_00,galfil_84,stfl_01,avu_98,review_07}.    Particularly,
S.Pagano et al.~\cite{pagano} for arbitrary junction lengths demonstrated the bifurcation curves of the fluxon
states in the three different regimes typically observed: small, intermediate, and long junction.  We consider
here the LJJ which barrier layer contains one resistive rectangular inhomogeneity, characterized by its width
$\Delta $, the  position of the center of the inhomogeneity $\zeta $ and the portion of Josephson current $k_J$
through it. The existence of the inhomogeneity leads to the local change of the Josephson current through the
junction, which can be modeled by $ j_C (x) = 1 + k_J$ inside of the inhomogeneity and $j_C (x) = 1$ outside of
it. At $k_J > 0$ the value of the current through the inhomogeneity exceeds the value of the current in other
parts of the junction and such inhomogeneity are considered as a shunt.~\cite{galfil_84} At $k_J \in [-1,0)$ the
value of the current through the inhomogeneity is less than  in other parts of the junction and such
inhomogeneity represents a microresistor. Changing the thickness of the barrier layer inside of the
inhomogeneity, we can model the transformation from the "shunt" to the "microresistor" inhomogeneity.  The value
$k_J = 0$ corresponds to a homogeneous junction.
\begin{figure}[!ht]
\onefigure[width=7cm]{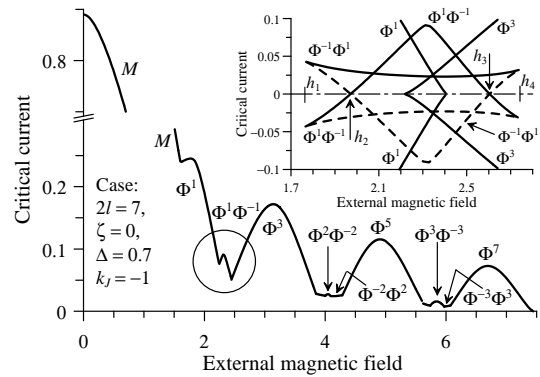} \caption{Critical curve of the LJJ with length $2l=7$ and width of inhomogeneity
$\Delta=0.7$. The insert shows the bifurcation curves in the marked by circle region.} \label{1}
\end{figure}
\begin{figure}[!ht]
\onefigure[width=7cm]{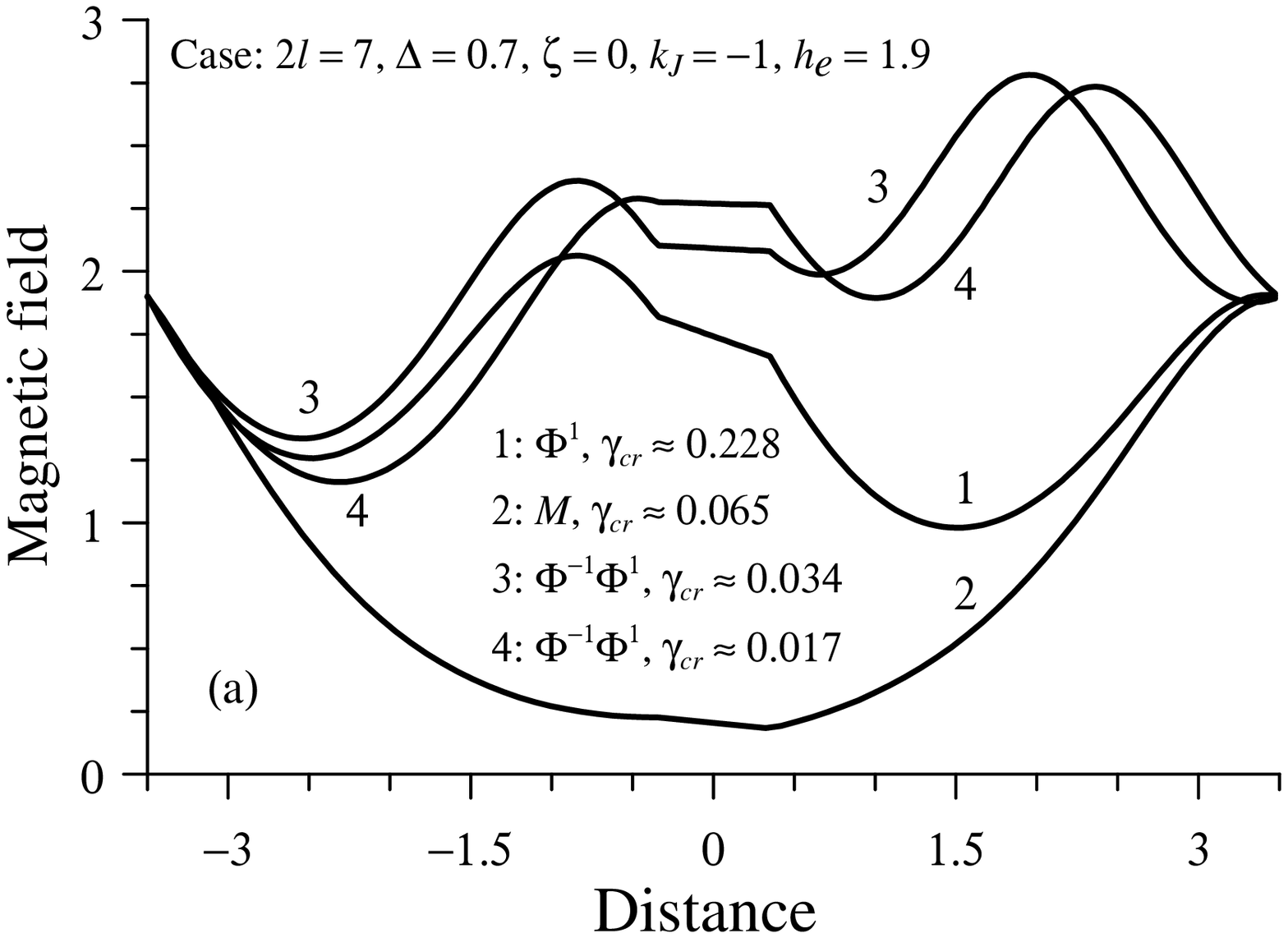}\onefigure[width=7cm]{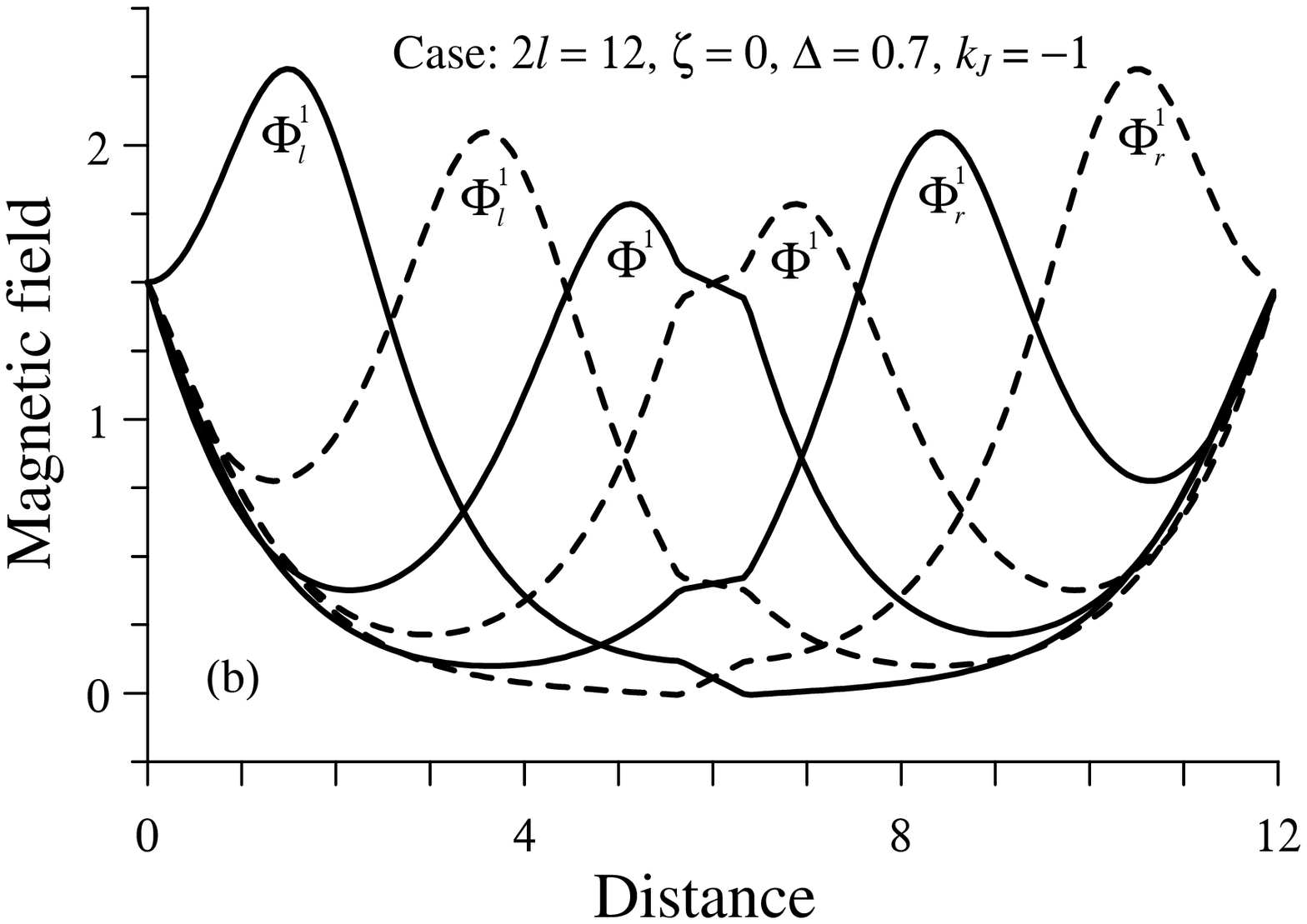} \caption{(a) The distribution of the magnetic field
$\varphi_x(x)$ along the junction for bifurcation states $M$, $\Phi^1$,  and $\Phi^{-1}\Phi^{1}$ in LJJ with
length $2l=7$, width of inhomogeneity $\Delta=0.7$ at $h_{e}=1.9$; (b) The distribution of the magnetic field
for LJJ with $2l=12$ for $\Phi^1$, $\Phi^1_l$ and $\Phi^1_r$ at $h_e=1.5$ and $\gamma = 0$.} \label{2}
\end{figure}
Results of numerical solution of non-linear eigenvalue problem ~\cite{review_07} eq.(~\ref{steq}),
eq.(~\ref{slp}) are presented in fig.~\ref{1}, where the curve ``critical current - external magnetic field'' of
the junction with $2l=7$ and the inhomogeneity with width $\Delta=0.7$ in the center of the junction is shown.
This curve is ``constructed'' as envelope of bifurcation curves eq.(~\ref{gamhe}), which belong to different
possible bound states ~\cite{review_07}.  The inhomogeneity leads to the appearance on the critical curves the
fragments of bifurcation curves of the stable mixed states like $\Phi^n\Phi^{-n}$ and $\Phi^{-n}\Phi^n$, $n =
1,2,\ldots$, which are not stable in homogeneous case. In one's turn this leads to the non-monotonic decrease of
maximums of critical curve, when the magnetic field increases. As we can see in fig.~\ref{1}, in the
investigated region of magnetic field the critical curve is an envelope of the critical curves for the Meissner
state $M$, mixed distributions $\Phi ^{1}\Phi ^{-1}$, $ \Phi^{-2}\Phi ^{2}$, $ \Phi^{-3}\Phi ^{3}$ and pure
fluxon states $\Phi^{1}$, $\Phi ^{3}$, $\Phi ^{5}$,  $\Phi ^{7}$. All ``even'' pure states like $\Phi^2$,
$\Phi^4$, etc., are unstable, so their bifurcation curves eq.(~\ref{gamhe}) do not share for critical curve of
junction as whole.

The insert to the fig.~\ref{1}  shows  the details of formation of the critical curve in the marked region. In
all our figures in this paper the solid lines correspond to the positive direction of current, the dashed lines
correspond to the negative direction of the current.  The critical curve of the junction consists of pieces of
the bifurcation curves for $\Phi^{1}$, $\Phi ^{1}\Phi^{-1}$ and $\Phi ^{3}$ states in this region.

The bifurcation curves for mixed $\Phi ^{1}\Phi^{-1}$ and $\Phi^{-1}\Phi ^{1}$ states demonstrate an interesting
peculiarity. As we can see, in the intervals of magnetic field $(h_{1},h_2)$ and $(h_3,h_{4})$ these CbC-states
are stable only if the current through the junction is not equal to zero.  A value of the current, which is
needed  for the creation of stable mixed distributions $\Phi^{1}\Phi^{-1}$ and $\Phi^{-1}\Phi^{1}$ in
CbC-regions,  depends on the value of the external magnetic field. The table presents the values of the physical
parameters for bifurcation states $\Phi^{-1}\Phi^{1}$ and $\Phi^{1}\Phi^{-1}$ at $h=1.9$ in comparison with  $M$
and $\Phi^1$ . Here the average value of the magnetic flux (number of fluxons) is defined as a functional
$N[\varphi] = \frac{1}{2\pi l}\int\limits_{-l}^{l} \varphi(x)\,dx$ and the full magnetic flux through the
junction is $\Delta\varphi = \varphi(l)-\varphi(-l)$.~\cite{review_07}
\begin{table}
\caption{Physical parameters for bifurcation states $M$, $\Phi^1$, $\Phi^{-1}\Phi^{1}$ and $\Phi^{1}\Phi^{-1}$}.
\label{tabl1}
\begin{center}
\begin{tabular}{|c|c|c|c|c|}
\hline
        Type & $\gamma_{cr}$ & N& $\Delta\varphi/2\pi$& $\varphi(0)/\pi$\\
        \hline
        $\Phi^1$          & 0.228   &  1.317     &  1.69  &   1.438\\
        $M$          & 0.065   &  0.073     &  0.847  &   0.048\\
        $\Phi^{-1}\Phi^{1}$ & 0.034   &  1.654 &  2.28   &   1.58\\
        $\Phi^{-1}\Phi^{1}$ & 0.017   &  1.43 &  2.19   &  1.329\\
        $\Phi^{1}\Phi^{-1}$ & $-0.017$ &  2.57 &  2.19   &  2.671\\
        $\Phi^{1}\Phi^{-1}$ & $-0.034$  &  2.346 &  2.28   &   2.42\\
        $M$          & $-0.065$ & $-0.073$ &  0.847  & $-0.048$\\
        $\Phi^1$           & $-0.228$ &  0.683     &  1.69   &   0.562\\
        \hline
\end{tabular}
\end{center}
\end{table}
For non-bifurcation pure fluxon  states at  $\gamma=0$ the number of vortices  equals to the integer number:
$N[\Phi^{\pm n}] = n$, while for non-bifurcation mixed fluxon states at $\gamma=0$ we obtain
$N[\Phi^n\Phi^{-n}]$ + $N[\Phi^{-n}\Phi^{n}] = 2n$.~\cite{review_07} For non-bifurcation stable Meissner
solution $N[M] = 0$. As we can see from the Table, the found bifurcation values satisfy the following relations.
For $\Phi^{n}$ states the half of sum of N for positive and negative currents equal to $n$. For CbC states we
should take the sum of $N[\Phi^n\Phi^{-n}]$  and $N[\Phi^{-n}\Phi^{n}]$ for opposite current directions to have
integer number $2n$. The magnetic flux in the center of junction $\varphi(0)/2\pi$ fulfil the similar
relationships.

The distribution of the internal magnetic field $\varphi_x(x)$ along the junction for bifurcation states $M$,
$\Phi^1$, $\Phi^{-1}\Phi^{1}$, and $\Phi^{-1}\Phi^{1}$ at the different values of $\gamma_{cr}$, presented in
the Table   is shown in fig.~\ref{2}(a) in case of LJJ with $2l=7$ and $\Delta = 0.7$ at $h=1.9$. The
inhomogeneity in LJJ attracts the bound states and it leads to the deformation of the distributions of the
magnetic field for these states.
\begin{figure}[!ht]
\onefigure[width=8cm]{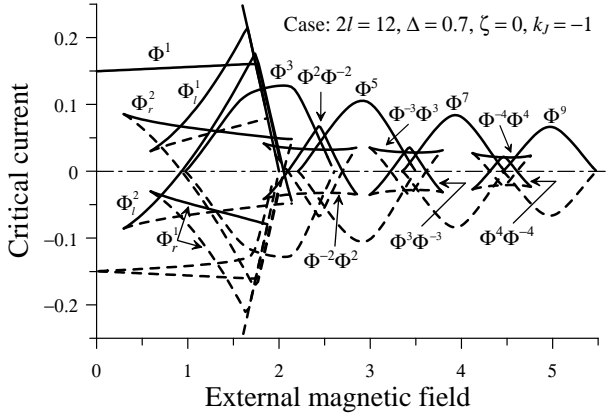} \caption{Bifurcation curves for LJJI at $2l = 12$ with the inhomogeneity in the
center with $\Delta=0.7$} \label{3}
\end{figure}
In fig.~\ref{3} we present  the bifurcation curves for vortices in LJJI with length $2l=12$ and the same  width
of inhomogeneity $\Delta=0.7$ in the center of junction and the zero amplitude of Josephson current through
inhomogeneity $j_{C}=0$. It shows clear the tendency of changing of the different bifurcation curves with
increase in magnetic field. The main features of this critical curve  coincide with the results for LJJ with
$2l=7$. Particularly, we observe here the mixed states as well, and as a result, the non-monotonic decrease of
maximums of the junction's critical curve with magnetic field.

But in contrast to the case $2l=7$, the new ``noncentral'' fluxon states are appeared here in addition to the
pinned to the inhomogeneity ``central'' ones. We found that in case under consideration the noncentral bound
states $\Phi^1_l$, $\Phi^1_r$, $\Phi^2_l$ and $\Phi^{-2}_r$  are stable only. The distribution of the internal
magnetic field along the junction for $\Phi^1$, $\Phi^1_l$ and $\Phi^1_r$ is shown in fig.~\ref{2}(b). The new
moment in case of $2l=12$ is a fact, that the states $\Phi^1_l$ and $\Phi^1_r$ are not stable without electric
current, at any value of external magnetic field. We have not found the stable $\Phi ^{1}\Phi^{-1}$ and
$\Phi^{-1}\Phi ^{1}$ states in this case also, so we consider that they are replaced with $\Phi^2_l$ and
$\Phi^{-2}_r$ when the length of LJJ is increased. Or we can say, that for LJJ with $2l=12$ the width's value
$\Delta = 0.7$ is small enough  to stabilize the $\Phi ^{1}\Phi ^{-1}$ and $\Phi ^{-1}\Phi ^{1}$ states. We can
see below in fig.~\ref{5}(a) the transformation of the $\Phi^2$ state into $\Phi ^{1}\Phi ^{-1}$ and $\Phi
^{-1}\Phi ^{1}$ with the increase in $\Delta$. The $\Phi^2_l$ and $\Phi^{-2}_r$ demonstrate the CbC-regions
which look like CbC-regions for mixed states $\Phi ^{1}\Phi ^{-1}$ and $\Phi ^{-1}\Phi ^{1}$ for small junction
length. Fig.~\ref{3} allows us to represent from the unified point of view the influence of the inhomogeneity on
the bifurcation curves of the fluxon states in LJJ. It leads to: a)mixed fluxon-antifluxon states $\Phi
^{-n}\Phi ^{n}$; b)non-monotonic decrease of maximums of the junction's critical curve with magnetic field; c)
right and left fluxon states; d)different position of the CbC-states of the different fluxon states.
\section{The influence of the  inhomogeneity width} 
\begin{figure}[!ht]
\onefigure[width=7cm]{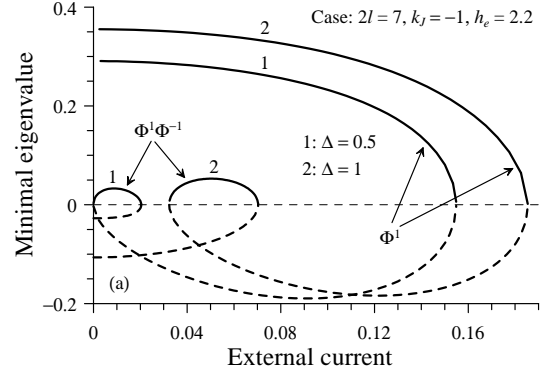} \caption{Dependence of the minimal eigenvalue $\lambda_{min}$ on the external
current $\gamma$ for $\Phi^1$ and $\Phi^1 \Phi^{-1}$ at two different values of the width of inhomogeneity
$\Delta = 0.1$ and $\Delta = 0.5$.} \label{4}
\end{figure}

The influence of the inhomogeneity width
 on the CbC-regions for mixed state $\Phi^1\Phi^{-1}$ in the case of $2l=7$ is demonstrated by
fig.~\ref{4}.  It shows  the dependence of the minimal eigenvalue $\lambda_{0}$ on the external current $\gamma$
for $\Phi^1$ and $\Phi^1\Phi^{-1}$ for the LJJ with two values of width of inhomogeneity $\Delta=0.5$ and
$\Delta=1$ at $h_e = 2.2$.  The zeroes of $\lambda_{0}$ determine the critical currents for corresponding
states. The two values of current $\gamma_{cr}$ determine the down and upper critical currents for the CbC
states at fixed value of the external magnetic field $h_{e}=2.2$. We can see that with increase in $\Delta$ the
interval of the CbC-region for $\Phi^1 \Phi^{-1}$ is increased.
\begin{figure}[!ht]
\onefigure[width=7cm]{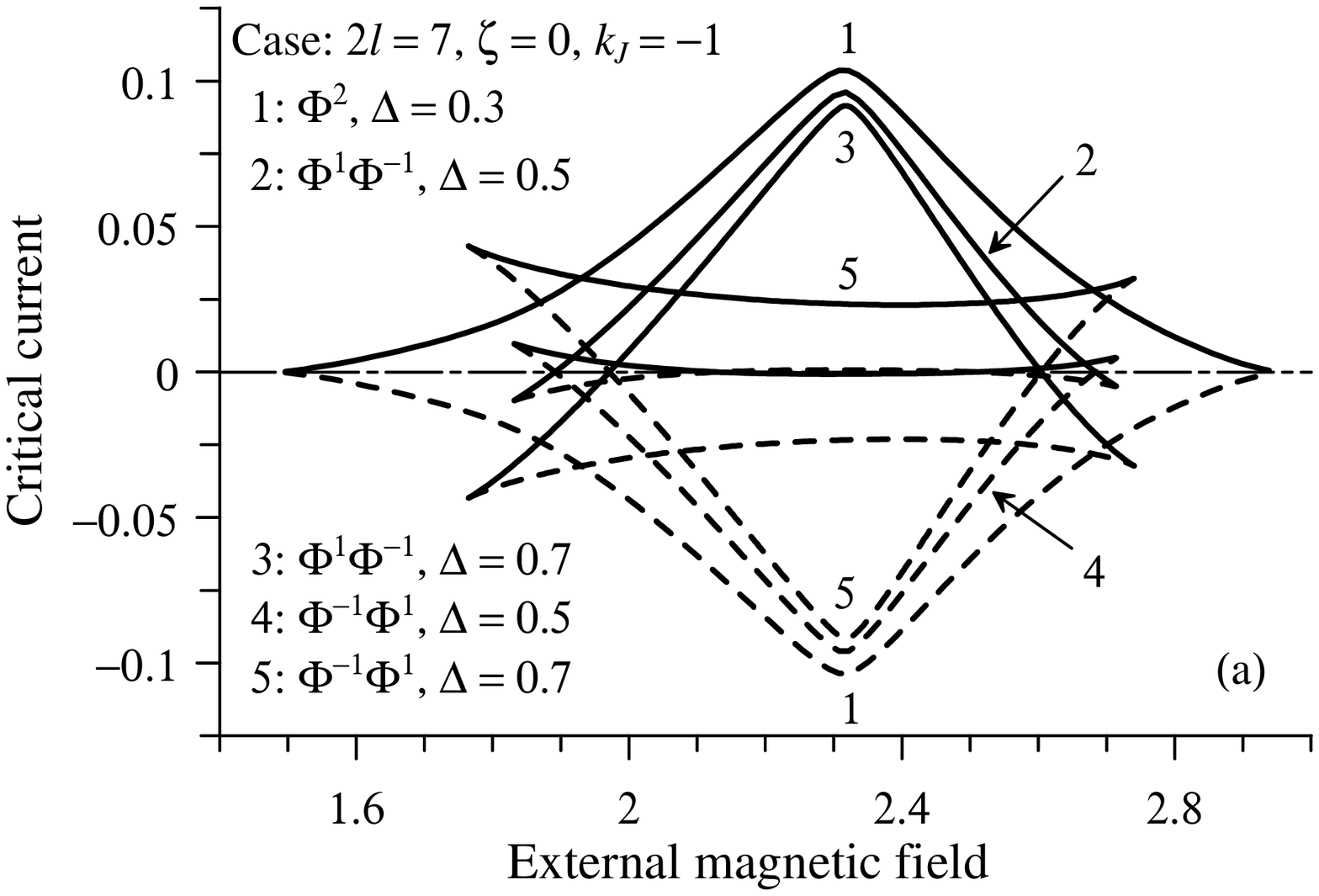}\onefigure[width=7cm]{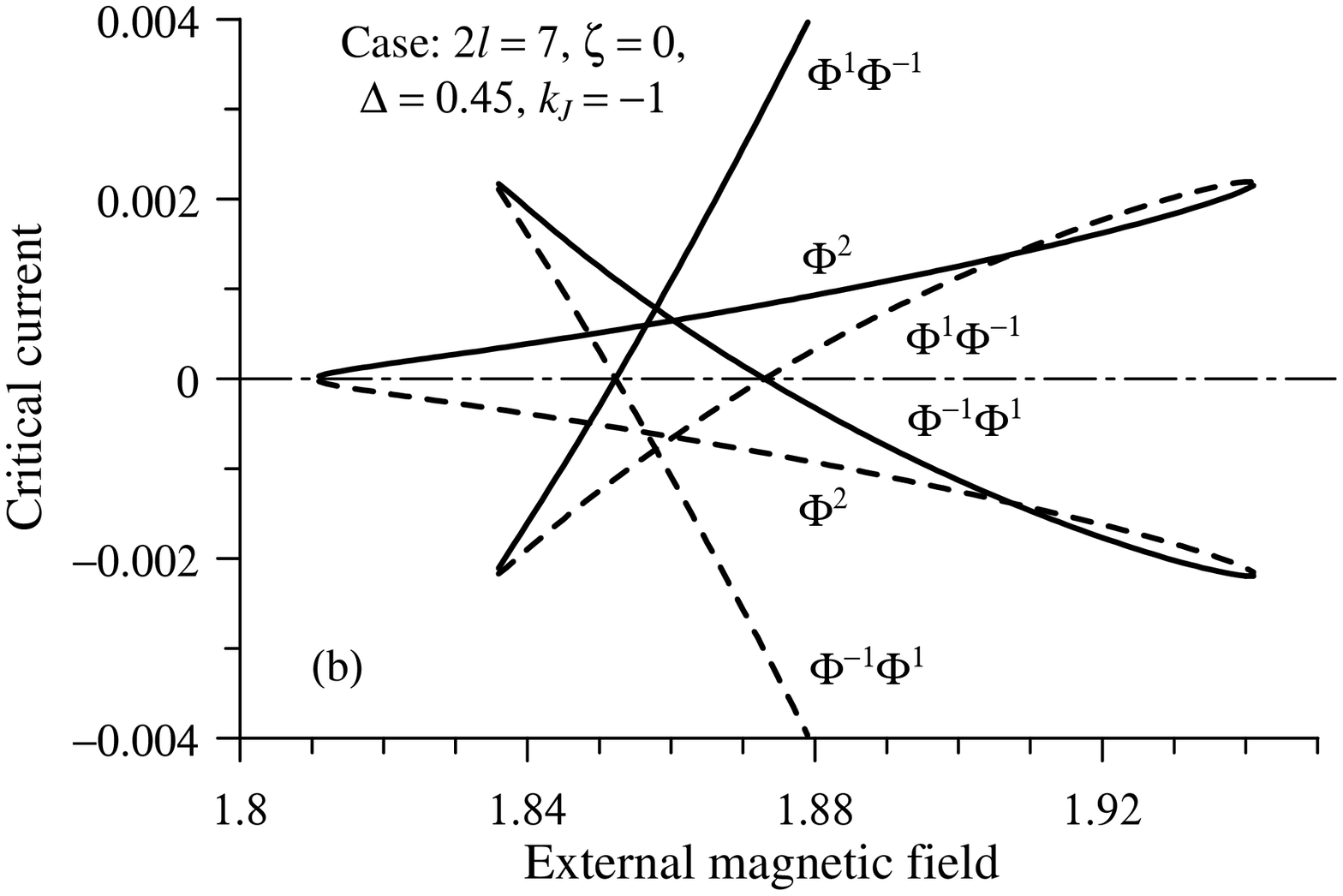}\onefigure[width=7cm]{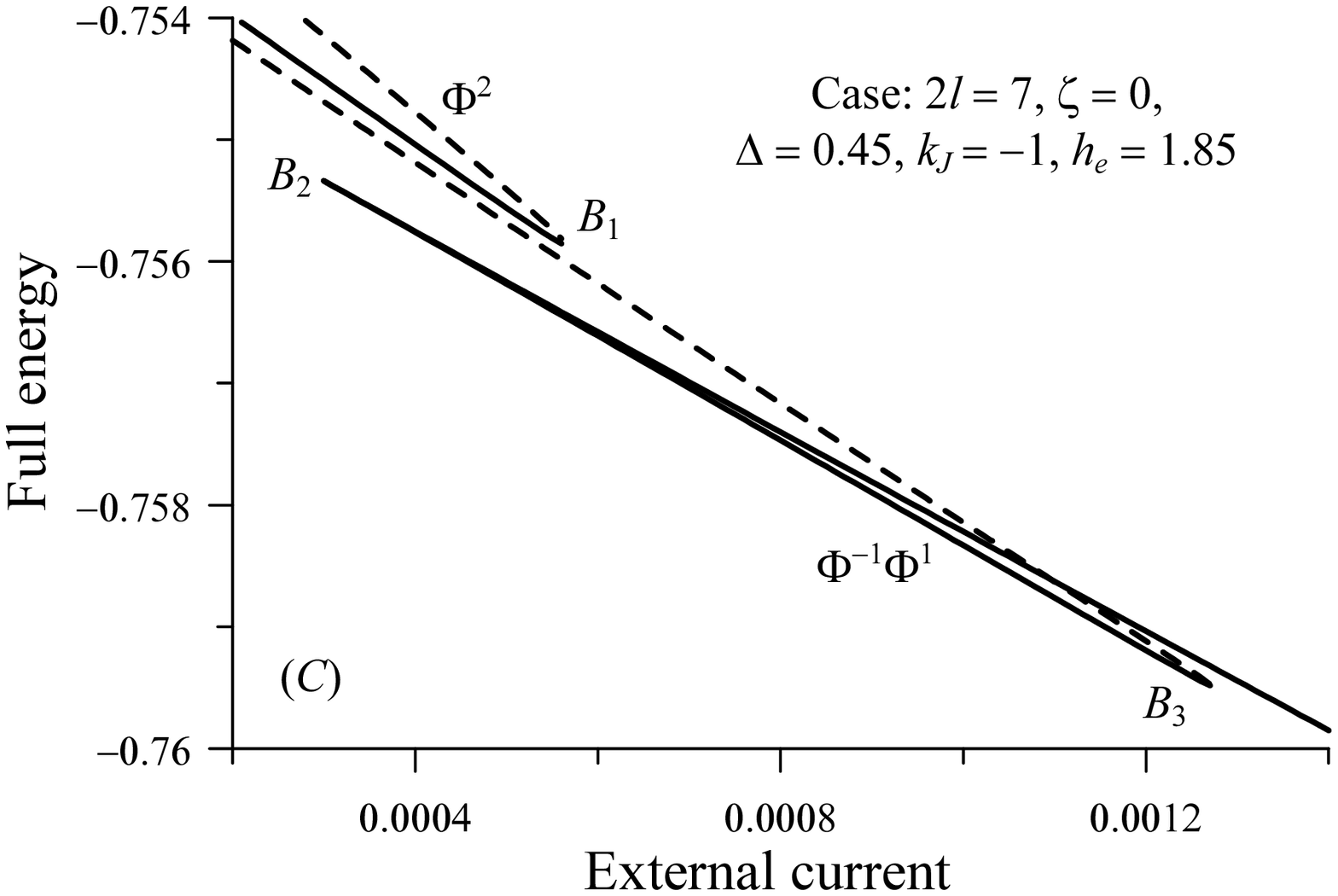} \caption{(a) The
transformation of the critical curve for $\Phi^2$ with the value of inhomogeneity's width; (b) The details of
the transformation of $\Phi^2$ at $\Delta = 0.45$; (c) The dependence of the total energy $F$  on external
current $\gamma$  for distributions $\Phi^2$, $\Phi^{-1}\Phi^1$  at $ h_e = 1.85$} \label{5}
\end{figure}
The influence of the width of the inhomogeneity $\Delta$ on the bifurcation curve for $\Phi^2$ is demonstrated
in fig.~\ref{5}(a). We show here the bifurcation curves at $\Delta = 0.3; 0.5; 1$. This figure demonstrates the
transformation $\Phi^2\rightarrow \left(\Phi^1\Phi^{-1}, \Phi ^{-1}\Phi ^{1}\right)$ with increase in $\Delta$.
There is a bifurcation region in the interval $0.3<\Delta<0.5$, where the state $\Phi^2$ is getting unstable,
but $\Phi^{1}\Phi ^{-1}$ and $\Phi ^{-1}\Phi ^{1}$  are stable. The details of such transformation are shown in
fig.~\ref{5}(b). As we can see, the state $\Phi^2$ is stable yet at $\Delta=0.45$, but the corresponding region
of magnetic field decreases with $\Delta$. This region disappears completely in the interval $0.45<\Delta<0.5$.

Actually, fig.~\ref{5} shows  in the current versus magnetic field plane typical features of the catastrophe
theory.~\cite{Poston}This correspondence is stressed by fig.~\ref{5}(c), where the dependence $F(\gamma)$ of the
total energy on $\gamma$ is shown for stable distributions $\Phi^2$, $\Phi^{-1}\Phi^1$ and related to them
unstable ones at $ h_e = 1.85$ (see fig.~\ref{5}b). The cusp in the $B_1$ point is a bifurcation point for $
\Phi^2$ with change in the external current. Correspondingly, the points $B_2$ and $B_3$ are the cusps of the
bifurcation points for mixed distribution $ \Phi^{-1}\Phi^1$. Note, that the interval along the $\gamma$-axis
which corresponds to the points $B_2$ and $B_3$ has not the point $\gamma = 0$, i.e. the current in $B_2$ point
is a "creation current", and in the point $B_3$ - "annihilation current" of the distribution $ \Phi^{-1}\Phi^1$.
The states the two horns (in the interval of magnetic field $h_1<h<h_2$ in the insert to fig.~\ref{1})  reflect
the fact that these states can not be stable without current through the junction. The detailed study of the
observed features in the correspondence of the catastrophe theory  will be done somewhere else.
\section{Shifting of the inhomogeneity}
\begin{figure}[!ht]
\onefigure[width=7cm]{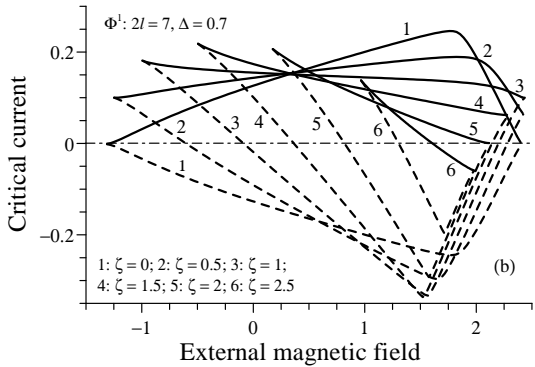} \caption{ Bifurcation curves for $\Phi^1$ at different position of the
inhomogeneity.} \label{4b}
\end{figure}
Fig.~\ref{4b} shows the bifurcation curves for $\Phi^1$ in LJJI with $2l = 7$, $\Delta = 0.7$ and $k_J = -1$ at
different position of the inhomogeneity $\zeta = 0; 0.5; 1; 1.5; 2$ and 2.5. We found that when  the
inhomogeneity is shifted from the center of the junction, the pure fluxon states have the CbC-regions as well.
The case $\zeta = 0$ correspond to the position of the inhomogeneity in the center of the junction.
\section{The influence of the current value through the inhomogeneity}
\begin{figure}[!ht]
\onefigure[width=7cm]{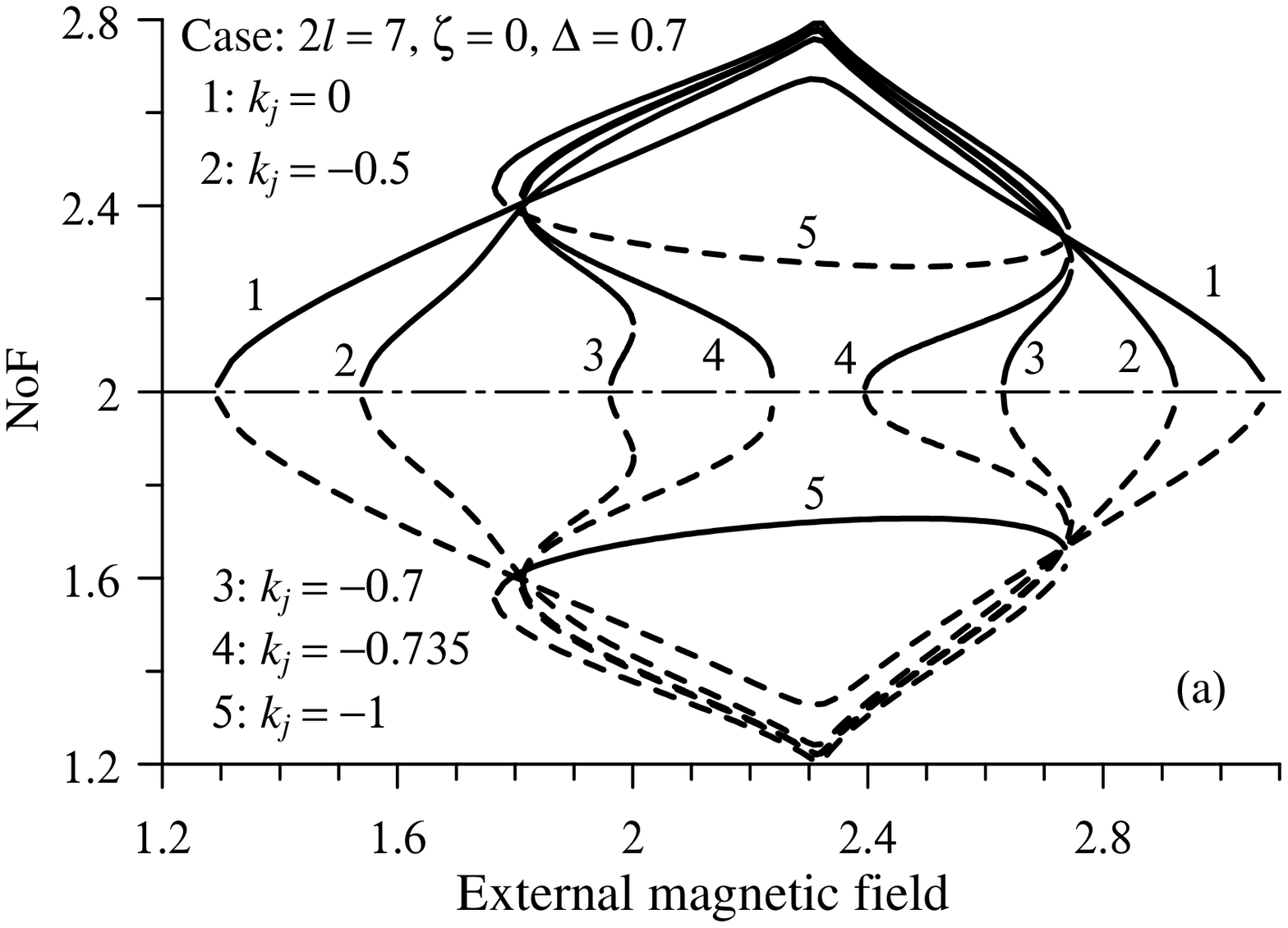}\onefigure[width=7cm]{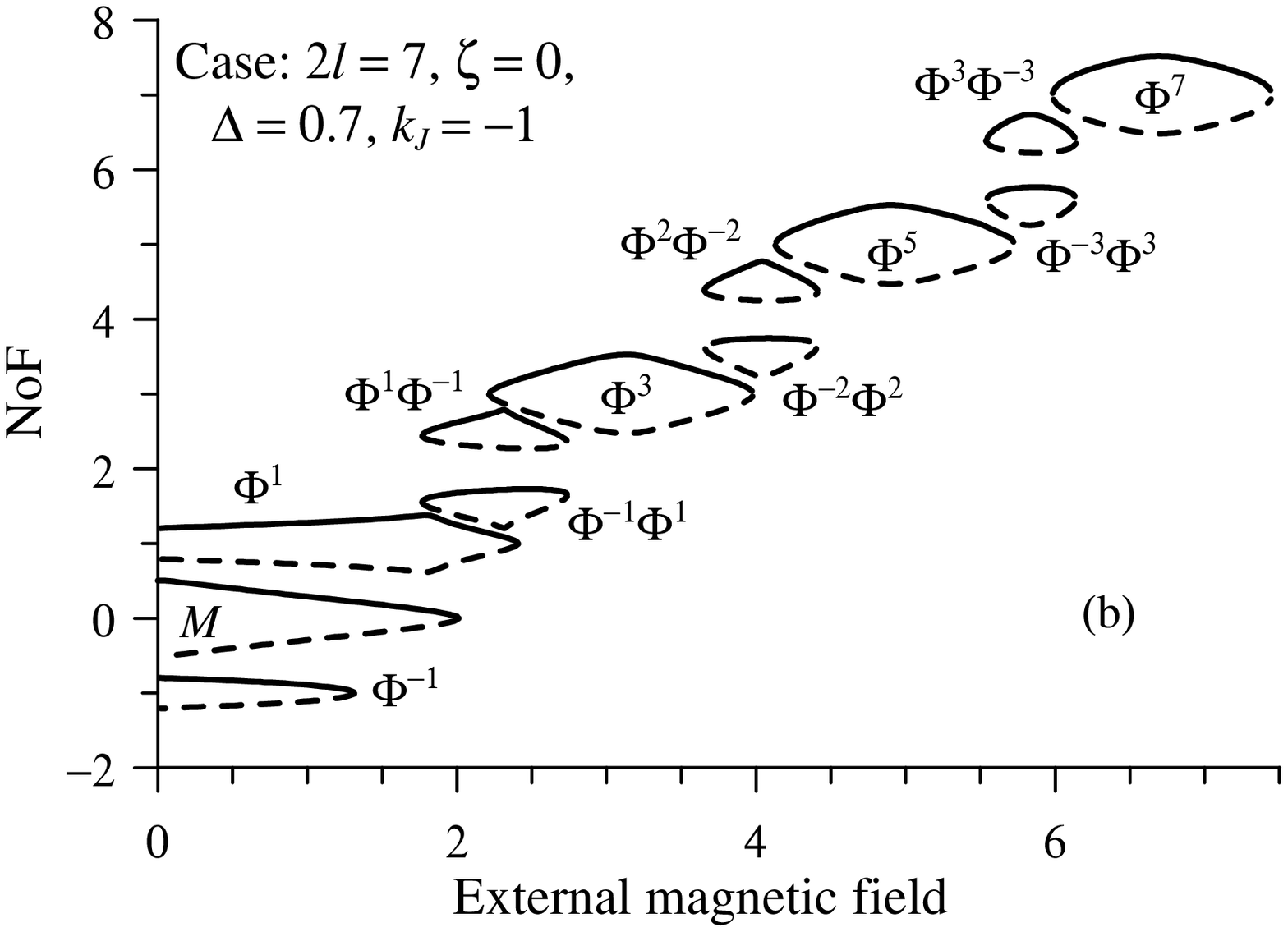} \caption{a) Transformation of the dependence
$N(h_{e})$  for $\Phi^2$ with  the amplitude of Josephson current through the inhomogeneity; (b) The dependence
$N(h_{e})$  in LJJ with $2l = 7$, $\Delta = 0.7$, $\zeta=0$ and $k_J = -1$.} \label{6}
\end{figure}

The decrease in the value of the
parameter $k_J$, which characterizes the amplitude of Josephson current through the inhomogeneity, leads to the
same transformation $\Phi^2\rightarrow \left(\Phi^1\Phi^{-1}, \Phi ^{-1}\Phi ^{1}\right)$ as we have observed
with increase in $\Delta$ . We demonstrate it for LJJI with $2l=7$  in fig.~\ref{6}(a), where the transformation
of the dependence $N(h_{e})$ for $\Phi^2$ with the value of the amplitude of Josephson current through the
inhomogeneity is shown. As we can see from this figure, a bifurcation of the states exists in the interval $-1 <
k_J < -0.735$.  We found that a new bifurcation point exists at $k_J\approx -0.77$. The dependence $N(h_{e})$ in
LJJ with $2l = 7$, $\Delta = 0.7$, $\zeta=0$ and $k_J = -1$ is shown in fig.~\ref{6}(b). It demonstrates clear
the relations for N for the pure and mixed bifurcation states we mentioned above. Particularly, for pure fluxon
state $\Phi^3$ sum of the $N(h_{e})$ for positive and negative current equals to 3, and for states
$\Phi^2\Phi^{-2}$, $\Phi^{-2}\Phi^{2}$ we get $N[\Phi^2\Phi^{-2}]$ + $N[\Phi^{-2}\Phi^{2}] = 4$.
\section{Exponentially shaped LJJ}
\begin{figure}[!ht]
\onefigure[width=7cm]{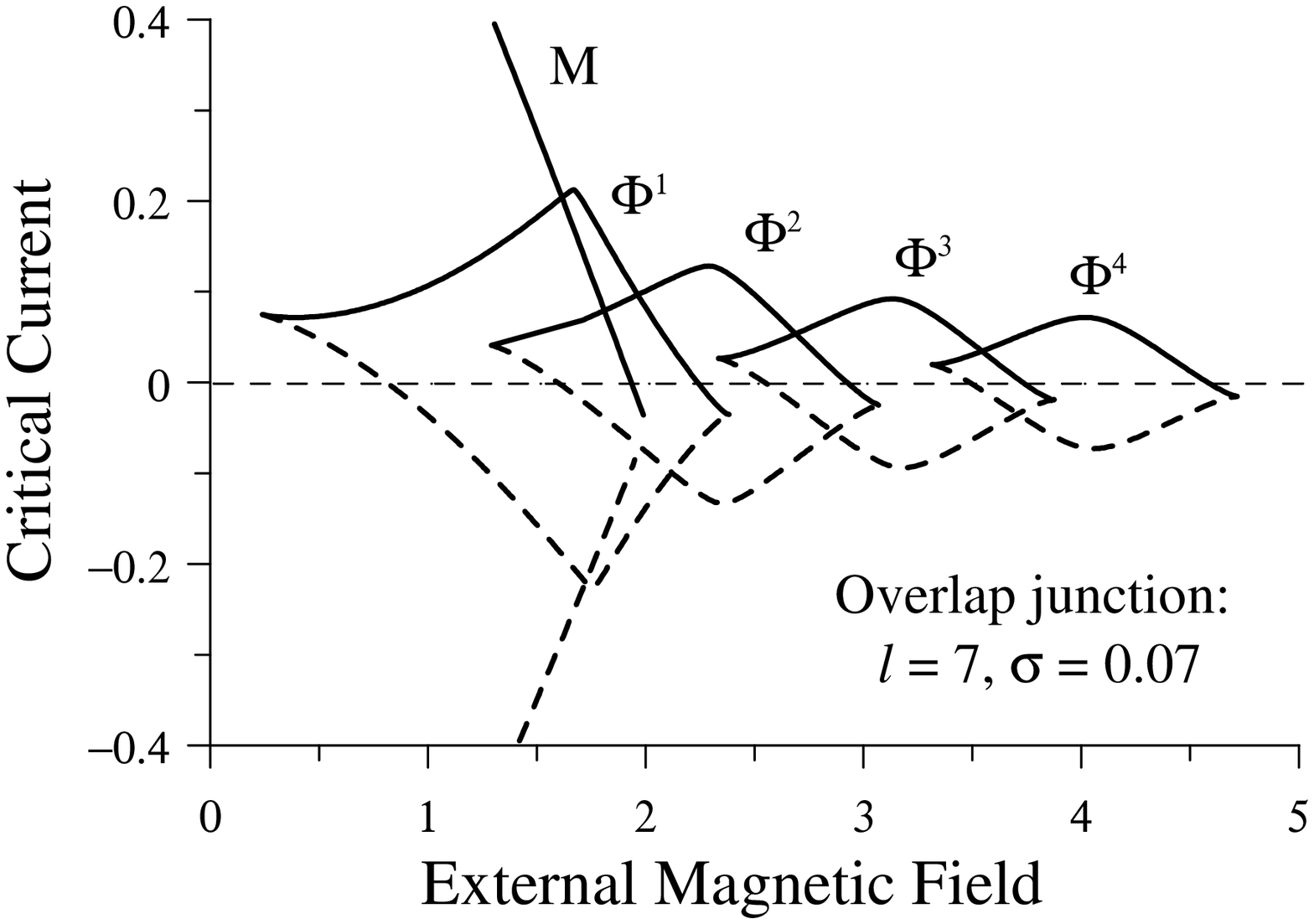}\onefigure[width=7cm]{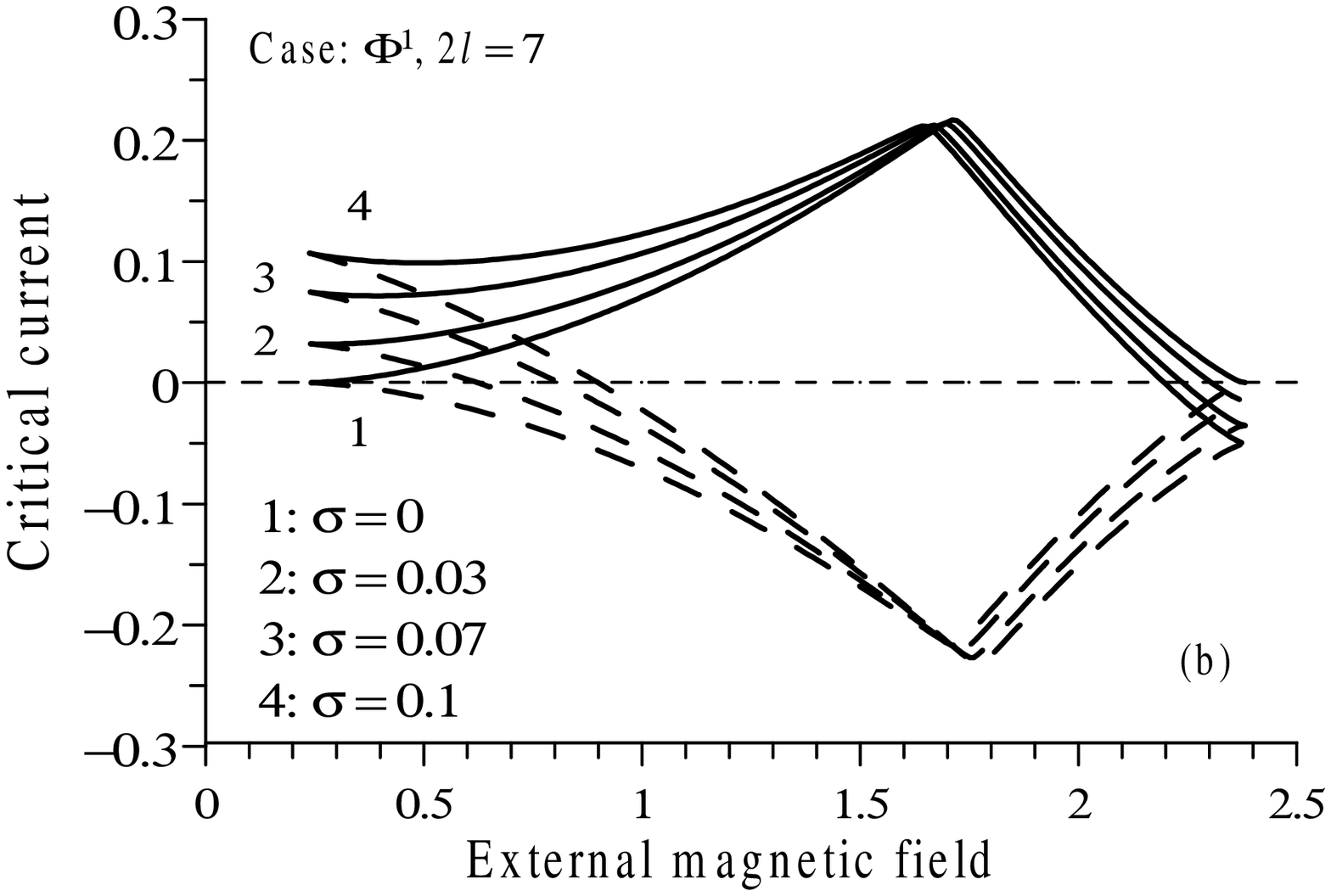} \caption{(a) Bifurcation curves for ESJJ at $2l=7$
and $\sigma =0.07$; (b) Bifurcation curves for $\Phi^1$ at different values of $\sigma$.} \label{7}
\end{figure}
 As a second  example, we consider an exponentially shaped LJJ
~\cite{sbs_04}-~\cite{sbs_05}. In Ref~{~\cite{sbs_05}} we demonstrated that the ESJJ is equivalent to the
Josephson junction with distributed inhomogeneity. So we may expect the CbC regions in the critical curves of
the ESJJ as well. The corresponding BVP for the ESJJ has the form:
\begin{equation}
\label{est} -\varphi_{xx} + \sin\varphi + \sigma(\varphi_x - h_{e}) - \gamma=0,
      \varphi_x(-l)=h_{e},\varphi_x(l)=h_{e}
\end{equation}
where $0\leq \sigma \ll 1$ is the form parameter.

Results of numerical solution of this BVP combined with the corresponding SLP ~\cite{sbs_04} are presented in
fig.~\ref{7}(a), where the bifurcation curves for Meissner and four first fluxon states are shown for $l=7$ and
$\sigma =0.07$. As we can see, the bifurcation curves of pure fluxon states in ESJJ demonstrate the CbC-regions.
The CbC-states are realized at positive current in low magnetic fields and negative current in high magnetic
fields. The exponential shape of the junction leads to the appearance of the distributed ``geometrical'' current
$\sigma\, (\varphi_x - h_{e})$, which modifies the bifurcation curves of the states in compare with the
rectangular homogeneous junction.~\cite{sbs_05} Bifurcation curves for $\Phi^1$  at $l=7$ and different values
of $\sigma$ are shown in fig.~\ref{7}(b). The increase in $\sigma$ leads to the increase of the CbC-region. With
increase in  $\sigma$ a situation might be realized, when the dependence of the upper critical current on the
external magnetic field  in CbC regions is getting non-monotonic.We found that such non-monotonic behavior of
the bifurcation curve $\Phi^1$ in low magnetic fields appears in ESJJ with increase in junction's length as
well.

In summary, we performed the numerical simulation of the critical curves of long Josephson junctions with
inhomogeneity and variable width. In both cases we demonstrate  the CbC-regions of magnetic field, where fluxon
states are stable only if the current through the junction is different from zero. We showed that the position
and size of these regions depend on length of the junction, its geometry, parameters of inhomogeneity and form
of the junction.  We consider that development of the experimental methods for detection of CbC-states will open
a perspective for their applications.\acknowledgments We thank I.V. Puzynin and N.M. Plakida for useful
discussion and cooperation. This work is partially supported by Sofia University Scientific foundation under
Grant No 135/2008 and  RFBR grant 08-02-00520-a.

\end{document}